# Vaporization and Layering of Alkanols at the Oil/Water Interface

*Aleksey M. Tikhonov[1] and Mark L. Schlossman[2]*

[1]University of Chicago, Center for Advanced Radiation Sources, and Brookhaven National Laboratory, National Synchrotron Light Source, Beamline X19C, Upton, NY 11973, tikhonov@bnl.gov

[2]University of Illinois at Chicago, Departments of Physics and Chemistry, 845 W. Taylor St., Chicago, IL 60607, schloss@uic.edu

## Abstract

This study of adsorption of normal alkanols at the oil/water interface with x-ray reflectivity and tensiometry demonstrates that the liquid to gas monolayer phase transition at the hexane/water interface is thermodynamically favorable only for long-chain alkanols. As the alkanol chain length is decreased, the change in excess interfacial entropy per area $\Delta S_a^\sigma$ decreases to zero. Systems with small values of $\Delta S_a^\sigma$ form multi-molecular layers at the interface instead of the monolayer formed by systems with much larger $\Delta S_a^\sigma$. Substitution of *n*-hexane by *n*-hexadecane significantly alters the interfacial structure for a given alkanol surfactant, but this substitution does not change fundamentally the phase transition behavior of the monolayers. These data show that the critical alkanol carbon number, at which the change in excess interfacial entropy per area decreases to zero, is approximately six carbons larger than the number of carbons in the alkane solvent molecules.

# I. INTRODUCTION

Adsorbed surfactant molecules at the liquid surface often form a monolayer and can be treated as a quasi-two-dimensional thermodynamic system [1]. For example, Langmuir monolayers of surfactant chain molecules at the surface of water have a complex phase diagram described by two thermodynamic parameters, *i.e.*, temperature $T$ and surface pressure $\Pi$ [1]. Over a wide range of surface concentrations the adsorbed chain molecules on the surface of water are in one of several solid monolayer phases whose symmetry is described by crystallographic simple point groups $C_n$ and $C_{nv}$. In contrast, soluble monolayers of the same surfactants at the hexane/water interface exhibit much simpler phase diagrams.

Long chain *n*-alkanols (CH$_3$(CH$_2$)$_{m-1}$OH, denoted C$_m$-alkanol) that are slightly soluble in *n*-hexane adsorb as monolayers at the *n*-hexane/water interface in a temperature range that is defined by the bulk concentration, $c$, of the alkanol in *n*-hexane and the ambient bulk pressure, $P$ [2-6]. According to our earlier study, monolayers of *n*-alkanols with 20, 22, 24, and 30 carbon atoms at the *n*-hexane/water interface are disordered (at $P$=1 atm), even at the lowest accessible temperatures (down to the temperature at which the alkanols precipitate from the bulk solution) [7, 8]. The structure normal to the interface of the C$_m$-alkanol monolayer in this low temperature phase can be described as consisting of two or three slabs, where each slab is characterized by its thickness and electron density. Typically, a slab corresponds to a section of the monolayer at a particular depth within the interface, such as the section occupied by the alkanol headgroups or tailgroups (Fig. 1). The headgroup slab is approximately 4Å thick with an electron density ~10% greater than that of water. An additional one or two slabs describe the progressive disordering of the chain from the –CH$_2$OH to the –CH$_3$ group. The second slab contains the part of the tailgroup chain closest to the headgroup. It is approximately 10 Å thick with an electron density similar to that of the rotator solid phases of bulk alkanes. The third slab contains the rest of the chain and consists of a disordered alkyl chain with significant conformational entropy. Its density is comparable to the density of liquid *n*-alkanes just above their melting temperature. A shorter chain *n*-alkanol, say with 20 carbons, can be adequately described by a two slab structure (slabs 1 and 3) that consists of the headgroup slab and the disordered chain slab.

When the temperature $T$ is increased the monolayer undergoes a phase transition at $T_o$ at which the interfacial density of the adsorbed molecules decreases considerably. The surfactant molecules leaving the interface are solvated in the bulk alkane. To a first approximation this phase transition is first order, representing vaporization of a quasi-two-dimensional liquid. However, for some materials, equilibrium coexistence of domains of the low and high-temperature monolayer phases was observed within some temperature range $\Delta T$ near $T_o$, where $\Delta T$ can be as large as tens of degrees [8, 9]. In these systems, it has been suggested that this phase transition can be explained as a second order transition determined by the competition of long-range and short-range interactions between the adsorbed dipolar surfactants [9-13].

In this paper we present evidence that sufficiently short alkanols do not undergo monolayer vaporization. As the alkanol chain length is decreased the interfacial excess entropy of monolayer vaporization goes to zero. As this critical chain length is approached the interface forms multi-molecular layers instead of monolayers. Evidence for these phenomena is obtained partially from a study of the thermodynamics of the phase transition at the oil/water interface in which the alkyl chain length is varied for both the alkane solvent and the alkanol surfactant. Additional evidence is obtained from x-ray reflectivity measurements that determine the electron density profile of $C_{24}$-alkanol and $C_{30}$-alkanol adsorbed layers at the *n*-hexane/water and *n*-hexadecane/water interfaces, as well as the profile of $C_{12}$-alkanol layers at the *n*-hexane/water interface.

## II. EXPERIMENTAL METHODS AND DATA

### A. Materials

$C_m$-alkanols ($CH_3(CH_2)_{m-1}OH$), *n*-hexadecane ($CH_3(CH_2)_{14}CH_3$), and *n*-hexane ($CH_3(CH_2)_4CH_3$) were purchased from Aldrich-Sigma. Alkanes were purified by passing them through activated alumina in a chromatography column. Alkanols (except $C_{12}$-alkanol) were re-crystallized twice at room temperature from an oversaturated *n*-hexane solution prepared by dissolving the alkanol in hexane at ~ 60 °C. $C_{12}$-alkanol (*n*-dodecanol, purity ~ 99%) was used as received. Purified de-ionized water was produced by a Barnstead NanoPure system.

The solubility of *n*-alkanol in *n*-alkane depends on the ratio, $r = m/m_o$, where *m* and $m_o$ are the numbers of carbons in the $C_m$-alkanol and the *n*-alkane solvent, respectively. The solubility of alkanols decreases significantly for large *r*. For example, the solubility of $C_{20}$-alkanol in *n*-hexane ($r = 3.33$) at $T = 300$ K is more than 20 times higher than it is for $C_{30}$-alkanol ($r = 5$). For convenience, we chose the concentrations of *n*-alkanols in the oil to adjust $T_o$ to be near room temperature.

## B. Interfacial Tension

The thermodynamic properties of the planar interfaces between bulk solutions of $C_m$-alkanols in hexane (or hexadecane) and bulk water were studied by measuring the interfacial tension, $\gamma$, using the Wilhelmy plate technique [8, 14]. At the phase transition temperature $T_o$ the interfacial tension curve, $\gamma(T)$, exhibits a sharp change in slope, which is associated with a change in interfacial excess entropy per unit area, $\Delta S_a^\sigma = \Delta(-d\gamma/dT)_{c,P}$ (Fig. 2). Throughout this paper, we denote a sample at a temperature below the phase transition as being in the low temperature phase of the interface. Likewise, the high temperature phase refers to the state of the sample interface above the phase transition. Fig 2 shows the temperature dependence, $\gamma(T)$, for monolayers of $C_{12}$- and $C_{30}$-alkanols at the hexane/water interface. Fig 3 shows the temperature dependence, $\gamma(T)$, for monolayers of $C_{24}$- and $C_{30}$-alkanols at the hexadecane/water interface. Fig. 4 shows the dependence of $\Delta S_a^\sigma$ vs. *r* for the solutions in hexane and hexadecane, where $r = m/m_o$ is the alkanol to alkane carbon number ratio.

According to earlier comprehensive studies of Aratono and co-authors [4, 15], $\Delta S_a^\sigma$ does not depend significantly on the concentration *c*. However, our data demonstrate that $\Delta S_a^\sigma$ depends strongly on *r*. For example, $\Delta S_a^\sigma$ for the $C_{30}$-alkanol is almost three times smaller at the hexadecane/water interface ($\Delta S_a^\sigma = 1.4 \pm 0.2$ mJm$^{-2}$K$^{-1}$) than at the hexane/water interface ($\Delta S_a^\sigma = 4.1 \pm 0.1$ mJm$^{-2}$K$^{-1}$). Also, $\Delta S_a^\sigma$ decreases when *r* is decreased at fixed solvent chain length.

For example, $\Delta S_a^\sigma$ for $C_{12}$-alkanol at the hexane/water interface is ten times smaller than it is for $C_{30}$-alkanol at the hexane/water interface.

## C. X-ray Reflectivity

X-ray reflectivity probes the electron density as a function of depth through the interface, but averaged over the in-plane interfacial region of the x-ray footprint [16]. The reflectivity data consist of measurements of the x-ray intensity reflected from the sample interface normalized by the incident intensity. These data are further modified by subtracting a background due primarily to scattering from the bulk liquids. The technique of x-ray reflectivity and its application to the study of liquid/liquid interfaces has been described previously in detail [8, 17-19]. We have used this technique to study molecular ordering and phase transitions in surfactant monolayers at oil/water interfaces [7-9, 13, 18, 20], the structure of neat oil/water interfaces [21-24], the adsorption of sodium ions at the oil/silica hydrosol interface [25], and the ordering of ions at the interface between two electrolyte solutions [26, 27].

Here, we use x-ray reflectivity to study the molecular ordering at planar interfaces between water and bulk solutions of alkanols in alkanes. The x-ray reflectivity data presented in this paper were obtained at beamline X19C of the National Synchrotron Light Source, Brookhaven National Laboratory. X-ray measurements were carried out on the same samples studied with tensiometry. A complete, detailed description of the experimental setup, as well as the temperature dependence of the reflectivity $R(q_z)$ for $C_{20}$-, $C_{22}$-, $C_{24}$-, and $C_{30}$-alkanol monolayers at the hexane/water interface was previously published [8, 17]. The x-ray wavelength was $\lambda = 0.825$ Å ($\Delta\lambda/\lambda \sim 2 \times 10^{-3}$).

Figures 5, 6, and 7 illustrate the x-ray reflectivity normalized by the Fresnel reflectivity, $R/R_F$ as a function of the wave vector transfer $q$, which has only one nonzero component, $q_z = (4\pi/\lambda)\sin\alpha$ (Fig. 1). The Fresnel reflectivity, $R_F$, is calculated for an ideal flat and smooth interface [16]. Figure 5a illustrates $R/R_F$ for the low temperature ($T$=21.9 °C) and high temperature ($T$=45.3 °C) phases of a $C_{24}$-alkanol monolayer at the hexane/water interface. Figure 5b illustrates $R/R_F$ for the low temperature ($T$=50.8 °C) and high temperature ($T$=81.9 °C) phases of $C_{24}$-alkanol

at the hexadecane/water interface. Figure 6a illustrates $R/R_F$ for the low temperature ($T$=24.5 °C) and high temperature ($T$=45.0 °C) phases of $C_{30}$-alkanol monolayer at the hexane/water interface. Figure 6b illustrates $R/R_F$ for the low temperature ($T$=24.9 °C) and high temperature ($T$=48.8 °C) phases of $C_{30}$-alkanol monolayer at the hexadecane/water interface. Finally, Fig. 7 illustrates $R/R_F$ for the low temperature ($T$= 8.0 °C) and high temperature ($T$= 55.0 °C) phases of $C_{12}$-alkanol at the hexane/water interface.

## III. INTERFACIAL MODELS

The reflectivity data in Figs. 5, 6, and 7 were analyzed using the first Born approximation that represents the reflectivity $R(q_z)$ as $R(q_z) = |F(q_z)|^2 R_F(q_z)$, where $F(q_z)$ is the structure factor of the surface, and $R_F(q_z)$ is the Fresnel reflectivity. The interfacial structure is represented by $L$ slabs of thickness $L_j$, and electron density $\rho_j$, where $j$ varies from 1 to $L$ (see Fig. 1). In addition, $L+1$ parameters determine the interfacial widths $\sigma_j$ between the slabs and the two bulk phases. The water surface is set to coincide with the $xy$ plane at $z = 0$. The interfacial electron density profile $\rho(z)$ is described by the following equation [28]:

$$\rho(z) = \frac{1}{2}(\rho_w - \rho_h) + \frac{1}{2}\sum_{j=0}^{L}(\rho_{j+1} - \rho_j) \, erf\left(\frac{t_j(z)}{\sigma_j \sqrt{2}}\right), \qquad (1)$$

where all electron densities are normalized to the value of bulk water such that $\rho_o = \rho_w \equiv 1$ and $\rho_{L+1} \equiv \rho_h$ is the normalized electron density of the bulk hexane, $t_j(z) = z + \sum_{i=0}^{j} z_i$, $z_i$ are the location of the interfaces as labeled in Fig. 1, and the error function is $erf(t) = (2/\sqrt{\pi})\int_0^t e^{-s^2} ds$. The model parameters are fit to the reflectivity data with a non-linear least squares fitting routine.

### A. High Temperature Phase

X-ray reflectivity at high-temperatures ($T \gg T_o$) for all systems can be described by a model with a single fitting parameter $\sigma$ that represents the effective interfacial width: $R(q_z) = R_F(q_z)\exp(-q_z^2\sigma^2)$. In this model, $\sigma^2 = \sigma_{cap}^2 + \sigma_{int}^2$, where the intrinsic width, $\sigma_{int}$, represents interfacial molecular ordering. The average interfacial width is increased by thermal fluctuations of the intrinsic structure [29, 30]. These fluctuations are represented by $\sigma_{cap}$, which is determined by the spectrum of capillary waves [31-33].

$$\sigma_{cap}^2 = \frac{k_B T}{2\pi\gamma} \ln\left(\frac{Q_{max}}{Q_{min}}\right), \qquad (2)$$

where $Q_{max} = 2\pi/a$ ($a \approx 5$ Å is of the order of the intermolecular distance), and $Q_{min} = q_z^{max} \Delta\beta/2$. The calculated value for $\sigma_{cap}$ is typically 3.5 Å to 4 Å.

For all samples with interfaces between water and a hexane solution of alkanols the interfacial width at high temperatures is significantly larger than it is at the neat hexane/water interface (for which it is $3.5 \pm 0.2$ Å). As an example, the measured width $\sigma$ for the sample with $C_{30}$-alkanol is $4.8 \pm 0.2$ Å with $\sigma_{int} = 2.9 \pm 0.5$ Å (see Table 1 for values for the other alkanols). The interfacial structure factor for the long chain $C_m$-alkanol monolayers (m ~ 12 to 30) in the gas monolayer phase is typically described by $\sigma_{int} \sim 3$ Å. Further resolution of this thin and low contrast intrinsic structure would require a significant increase in the range of wave vector transfer. However, the reflectivity from the water interface with a hexadecane solution of alkanols at high temperature can be described by an interfacial width $\sigma$ that is within statistical error of the calculated capillary wave value (Table 1).

## B. $C_{24}$-alkanol Low Temperature Phase

Comparison of panels (a) and (b) in Fig. 5 shows that the structure factors of $C_{24}$-alkanol at the low temperature hexane/water and hexadecane/water interfaces are different. The structure of $C_{24}$-alkanol at the hexane/water interface is that of a monolayer of molecules with a partially disordered tailgroup. Approximately half of the tailgroup near the terminal methyl has an electron density that corresponds to an alkane liquid. The half of the tailgroup closer to the headgroup is more ordered, with an electron density similar to that of solid rotator alkane phases. As we previously showed, the headgroup region is about 10% denser than expected, which indicates that water molecules can penetrate the region (*i.e.*, the same interfacial depth) occupied by the alkanol headgroups [7, 8].

The reflectivity from $C_{24}$-alkanol at the hexane/water interface is described well by a three slab model of a monolayer (see also Table 1) [8]. However, the reflectivity from $C_{24}$-alkanol at the hexadecane/water interface cannot be described well by either a two or three slab model of a monolayer. The primary difficulty in fitting the reflectivity from $C_{24}$-alkanol at the hexadecane/water interface is the broad first peak that extends to low values of $q_z$. This broad peak is properly described by two peaks. The peak at lower $q_z$ reveals that the interfacial structure is thicker than a monolayer of molecules. A good fitting of these data results from a model that represents a bilayer of molecules, though three slabs of electron density are required for this representation. The model profiles of electron density for the interfacial structure of $C_{24}$-alkanol at these two interfaces can be compared in Fig. 8. The structure at the hexadecane/water interface is twice as thick as the monolayer at the hexane/water interface. Table 1 and Fig. 8 demonstrate that the first two slabs of the electron density profile at the hexadecane/water interface correspond to a layer of molecules with a normalized density of essentially 0.97, similar to that of bulk alkane rotator phases. Note that the large density of slab 2 is coupled to a small slab thickness, therefore, the final profile illustrated in Fig. 8 has only a very small increase in the density in this region. Slab 3 has a lower normalized density, 0.74, that corresponds to a disordered layer of molecules. Slab 3 corresponds to the second layer of molecules. Some of the parameters in the model of the hexadecane/water interface have large error bars because they are correlated with other parameters, however, the electron density profile shown in Fig. 8 is essentially unchanged for good fits.

The electron density profile allows us to calculate $N$, the number of electrons per area of the interface (see Table 1) by integrating just the monolayer part of the profile over the distance normal to the interface (equivalently, $N = \rho_{\text{water}} \sum_i \rho_i L_i$, where $\rho_{\text{water}}$ is the absolute electron density of water). Dividing the number of electrons per alkanol by $N$ yields the area per alkanol molecule $A$ if one assumes that only alkanol molecules exist within the interfacial region. The average $A$ in the $C_{24}$-alkanol bilayer of molecules is $23 \pm 1$ Å$^2$, which is comparable to $A = 22.4 \pm 1$ Å$^2$/molecule for the $C_{24}$-alkanol monolayer at the hexane/water interface.

## C. $C_{30}$-alkanol Low Temperature Phase

Comparison of panels (a) and (b) in Fig. 6 shows that the monolayer of $C_{30}$-alkanol at the hexane/water interface appears to be approximately 25% thicker than at the hexadecane/water interface because the period of oscillations in $R/R_F$ at the hexane/water interface is smaller. The reflectivity from $C_{30}$-alkanol at both interfaces represents a monolayer of molecules. Three slabs are required to represent this monolayer at the hexane/water interface, but two or three slabs can be used to represent the monolayer at the hexadecane/water interface.

At the hexane/water interface the structure of the $C_{30}$-alkanol monolayer is very similar to the structure of the $C_{24}$-alkanol monolayer at the hexane/water interface. The parameters listed in Table 1 are almost identical for the two monolayers except that slabs 2 and 3, which describe the alkyl chains, are thicker as expected for the longer molecule.

The two slab model provides the best fit to $R/R_F$ from $C_{30}$-alkanol at the hexadecane/water interface. The normalized electron density of 0.89 in the second slab is slightly lower than the density in solid alkane rotator phases, which are the lowest density solid bulk alkane phases and have a normalized density of 0.92 – 0.96. The density of slab 2 is also slightly higher than electron densities of bulk liquid alkane phases. If the alkyl tailgroups were organized similar to alkanes in a rotator phase, then the tilt of the molecules would be $\theta \approx 48$ deg, where $\theta$ is the angle between the normal to the interface and the molecular axis ($\cos\theta = (L_1 + L_3)/L_{trans}$, where $L_{trans} \approx 42$ Å is the length of the all-trans $C_{30}$-alkanol molecule). To our knowledge, this angle is much larger than any previously observed for solid monolayers of this type of molecule at the liquid/liquid or liquid/vapor interface [1]. Both the values of electron density and the calculated tilt angle argue against the presence of ordered alkyl chains and demonstrate that the two slab model describes a disordered monolayer. The electron density parameters indicate that the $C_{30}$-alkanol monolayer is more disordered at the hexadecane/water interface than at the hexane/water interface.

The two slab model may not adequately represent the structure of the monolayer because of the nearly perfect contrast matching that might be expected between a liquid-like ordering of the tailgroup with bulk hexadecane. An example of this liquid-like ordering is given by the electron density of slab 3, with normalized density 0.79, of the $C_{30}$-alkanol at the hexane/water interface. The possibility of contrast matching is apparent when this density, 0.79, is compared to the normalized density of bulk hexadecane, which is 0.80 at room temperature. This indicates that the x-ray reflectivity measurements may not be sensitive to the end of the tailgroup and may explain why the $C_{30}$-alkanol monolayer at the hexadecane/water interface appears to be thinner than at the hexane/water interface.

The contrast matching, along with the experimental range of wave vector transfer and the good quality of the two slab fit, prevents us from specifying unique model parameters for a three slab fit. In Table 1 we present one possible three slab fit, which was determined by fixing the interfacial width to the capillary wave value and fixing the thickness parameter of the third slab to have the same value as in the three slab fit of $C_{30}$-alkanol at the hexane/water interface. This fit determines the electron density in slab 2 to be lower than that determined for $C_{30}$-alkanol at the hexane/water interface. Therefore, the $C_{30}$-alkanol at the hexadecane/water interface is more disordered than at the hexane/water interface, consistent with the two slab fit.

The model electron density profiles for the low temperature phase of $C_{30}$-alkanol at the hexane/water and hexadecane/water interfaces can be compared in Fig. 9. The monolayer at the hexadecane/water interface has lower density than at the hexane/water interface. The three slab model for $C_{30}$-alkanol at the hexadecane/water interface , as compared to the two slab model, has a slightly larger number of electrons per unit area. This suggests that part of the $C_{30}$-alkanol monolayer is hidden by contrast matching. The area per molecule of $28 \pm 4$ Å$^2$ (two slab model) or $25$ +0.3/-4 Å$^2$ (three slab model) is slightly larger than the $23.6 \pm 1$ Å$^2$ per $C_{30}$-alkanol at the hexane/water interface, though the statistical errors are large on the fits for $C_{30}$-alkanol at the hexadecane/water interface. These results are consistent with a more disordered monolayer at the water/hexadecane interface.

### D. $C_{12}$-alkanol Low Temperature Phase

The x-ray reflectivity data from $C_{12}$-alkanol adsorbed to the hexane/water interface at a temperature (8.0 °C) well below the phase transition is shown in Fig. 7. A measurement at 20.6 °C was similar (not shown). The contrast between $C_{12}$-alkanol and hexane is low and the reflectivity peaks are weak. The position of this peak at low $q_z$ demonstrates that the adsorbed film is thicker than a monolayer. The lack of any other prominent peaks indicates that the adsorbed film cannot be modeled by a single electron density slab. A minimum of three electron density slabs are required to model the position of the low $q_z$ peak and to closely approximate the rest of the reflectivity data. The parameters shown in Table 1 indicate that the thickness of each slab is approximately the length of an all-trans $C_{12}$-alkanol molecule, therefore the adsorbed film consists of three layers of molecules. This analysis does not rule out the presence of a larger number of molecular layers. In fact, a four layer fit provides a slightly better fit to the data, but we have chosen to present the three layer fit because it represents the minimum number of layers required for an acceptable fit. The parameters in Table 1 and Fig. 10 demonstrate that the electron density is smaller for layers further from the water surface.

Additional evidence for the multi-molecular layer adsorption is provided by the total number of electrons per area $N$ in the electron density profile. The value of 16 e$^-$/Å$^2$ far exceeds the number of electrons per area in the most closely packed monolayer of $C_{12}$-alkanol. If we assume that only $C_{12}$-alkanol molecules are in the interfacial region and that there are three layers of $C_{12}$-alkanol molecules, then the area per molecule is 20 Å$^2$. This value is typical of packing of alkanes in a solid rotator phase. However, the electron densities of slabs 2 and 3 are too small to represent layers of close packed all-trans $C_{12}$-alkanol molecules. These observations are consistent if the layers contain other molecules, most likely hexane, mixed with the $C_{12}$-alkanol molecules. If other molecules were mixed into the layers, then the area per alkanol molecule would increase. For example, if there were a hexane molecule for every three $C_{12}$-alkanol molecules at the interface, then the average area per $C_{12}$-alkanol molecule would be 23 Å$^2$, which represents liquid packing of the chains.

## IV. DISCUSSION

We have used x-ray reflectivity and interfacial tension measurements to probe the molecular ordering at the interface between water and both hexane and hexadecane solutions of $C_m$-alkanols (with m = 12, 24, and 30). These data reveal two important features of surfactant ordering at the alkane/water (liquid/liquid) interface. First, there is a very strong dependence of the structure of the adsorbed layer on the length of the alkane used for the solvent. Second, the nature of the adsorption, or vaporization, transition, changes dramatically when the alkanol chain is only six to eight carbons longer than the solvent alkane chain.

Our earlier measurements had shown that alkanol monolayers at the hexane/water interface undergo a vaporization phase transition as a function of temperature from a condensed liquid monolayer at low temperatures to a dilute gas monolayer at high temperatures [7, 8]. These observations corresponded to the larger values of $r$ shown in Fig. 4 for which the alkanol carbon number (20 to 30) far exceeded the alkane (hexane) carbon number (6). For these systems, our interfacial tension data revealed a large change in interfacial excess entropy $\Delta S_a^\sigma$ across this transition. However, Fig. 4 demonstrates that increasing the length of the alkane solvent to 16 carbons (hexadecane) significantly decreases $\Delta S_a^\sigma$ for $C_{24}$-alkanol and $C_{30}$-alkanol.

The interfacial excess entropy $S_a^\sigma$ represents the difference in entropy between a molecule in the bulk phase and one at the interface. Therefore, a significant reduction in $\Delta S_a^\sigma$ for say, $C_{30}$-alkanol in hexadecane/water as compared to hexane/water, indicates that $C_{30}$-alkanol is either more ordered in the bulk hexadecane or less ordered in the low temperature interfacial phase at the hexadecane/water interface or, possibly, both. One might expect some small difference in ordering between $C_{30}$-alkanol in bulk hexadecane and in bulk hexane, because bulk hexadecane is closer to its freezing point (18 °C) than hexane. However, the primary difference in ordering is at the interface, as revealed by x-ray reflectivity. $C_{30}$-alkanol monolayers are more disordered in the low temperature phase at the hexadecane/water interface than at the hexane/water interface. This demonstrates a strong dependence of the interfacial ordering on the molecular length of the alkane solvent.

As the alkanol chain length is reduced $\Delta S_a^\sigma$ approaches zero (Fig. 4). Extrapolation of the curves in Fig. 4 indicates that $\Delta S_a^\sigma = 0$ will occur for alkanol chains approximately 6 carbons longer than the alkane solvent chains. Two of the systems studied, $C_{12}$-alkanol at the hexane/water interface and $C_{24}$-alkanol at the hexadecane/water interface exhibited a small, though apparently nonzero, $\Delta S_a^\sigma$. These two systems had a remarkable interfacial structure consisting of a tri-molecular layer for $C_{12}$-alkanol and a bi-molecular layer for $C_{24}$-alkanol. These phenomena suggest the presence of a wetting transition.

Insight into the phenomena discussed here can be obtained by considering the adsorption of a single component gas onto a solid substrate. It is well known that gas adsorption can yield a single layer or multiple layers of molecules on a solid substrate, depending upon the thermodynamic conditions (for a review, see [34, 35]). If the one-component gas phase is kept at a fixed temperature, then adsorption on the solid substrate will increase as the pressure (or, alternatively, the chemical potential) is varied to bring the bulk gas phase closer to bulk liquid-gas coexistence. In the oil (with alkanol surfactant)/water system, the analog of the gas phase is the dilute alkanol solution in alkane oil and the analog of the solid substrate is the water phase. Ideal solution theory expresses the chemical potential of a dilute solution of molecules in a mathematical form very similar to the chemical potential of an ideal gas. The solvent acts to renormalize the interactions between the solvated molecules. However, if the solvent is identical to the solvated molecules, then the solution is just a single component liquid. This suggests that by varying the molecular length of the alkane solvent, the effective interaction between the alkanol molecules is changed from gas-like (for alkanes much shorter than the alkanols) to liquid-like (for alkanes of nearly the same length as the alkanols). Increasing the molecular length of the alkane solvent in the oil/water system is roughly analogous to approaching liquid-gas coexistence in the one-component gas/solid substrate system.

Our experimental results suggest that the liquid alkane/water interface is wet by alkanol layers as $\Delta S_a^\sigma$ approaches zero. The largest number of adsorbed layers was observed for $C_{12}$-alkanol, which has the smallest $\Delta S_a^\sigma$. For systems with even smaller $\Delta S_a^\sigma$, it is possible that the number of adsorbed layers increases. We plan to carry out x-ray reflectivity measurements on such systems to explore their adsorption behavior.

**ACKNOWLEDGMENTS**

Use of the National Synchrotron Light Source, Brookhaven National Laboratory, was supported by the U.S. Department of Energy, Office of Science, Office of Basic Energy Sciences, under Contract No. DE-AC02-98CH10886. Beamline X19C at NSLS is supported by funding from the University of Illinois at Chicago, ChemMatCARS at the University of Chicago, and SUNY Stony Brook.

**References**

[1]     Kaganer V M, Mohwald H and Dutta P 1999 *Rev. Mod. Phys.* **71** 779

[2]     Jasper J J and Houseman B L 1963 *J. Phys. Chem.* **67** 1548

[3]     Matubayasi N, Motomura K, Aratono M and Matuura R 1978 *Bull. Chem. Soc. Jpn.* **51** 2800

[4]     Motomura K, Matubayasi N, Aratono M and Matuura R 1978 *J. Coll. Int. Sci.* **64** 356

[5]     Lin M, Firpo J-L, Mansoura P and Baret J F 1979 *J. Chem. Phys.* **71** 2202

[6]     Ikenaga T, Matubayasi N, Aratono M, Motomura K and Matuura R 1980 *Chem. Soc. Jap.* **53** 653

[7]     Tikhonov A M and Schlossman M L 2003 *J. Phys. Chem. B* **107** 3344

[8]     Tikhonov A M, Pingali S V and Schlossman M L 2004 *J. Chem. Phys.* **120** 11822

[9]     Li M, Tikhonov A and Schlossman M L 2002 *Europhys. Lett.* **58** 80

[10]    McConnell H M 1991 *Annu. Rev. Phys. Chem.* **42** 171

[11]    Mohwald H 1990 *Annu. Rev. Phys. Chem.* **41** 441

[12]    Marchenko V I 1986 *JETP* **63** 1315

[13]    Tikhonov A M, Li M, Mitrinovic D M and Schlossman M L 2001 *J. Phys. Chem. B* **105** 8065

[14]    Adamson A W, Physical Chemistry of Surfaces, 5th ed., John Wiley & Sons, New York, NY, 1990

[15]    Takiue T, Uemura A, Ikeda N, Motomura K and Aratono M 1998 *J. Phys. Chem.* **102** 3724

[16]    Als-Nielsen J and McMorrow D, Elements of Modern X-ray Physics, John Wiley & Sons, Inc., Hoboken, 2001

[17]    Schlossman M L, Synal D, Guan Y, Meron M, Shea-McCarthy G, Huang Z, Acero A, Williams S M, Rice S A and Viccaro P J 1997 *Rev. Sci. Instrum.* **68** 4372

[18]    Zhang Z, Mitrinovic D M, Williams S M, Huang Z and Schlossman M L 1999 *J. Chem. Phys.* **110** 7421

[19]    Mitrinovic D M, Williams S M and Schlossman M L 2001 *Phys. Rev. E* **63** 021601

[20]  Pingali S V, Takiue T, Luo G, Tikhonov A M, Ikeda N, Aratono M and Schlossman M L 2005 *J. Phys. Chem. B* **109** 1210

[21]  Mitrinovic D M, Tikhonov A M, Li M, Huang Z and Schlossman M L 2000 *Phys. Rev. Lett.* **85** 582

[22]  Tikhonov A M, Mitrinovic D M, Li M, Huang Z and Schlossman M L 2000 *J. Phys. Chem. B* **104** 6336

[23]  Luo G, Malkova S, Pingali S V, Schultz D G, Lin B, Meron M, Graber T, Gebhardt J, Vanysek P and Schlossman M L 2005 *Electrochem. Comm.* **7** 627

[24]  Luo G, Malkova S, Pingali S V, Schultz D G, Lin B, Meron M, Benjamin I, Vanysek P and Schlossman M L 2006 *J. Phys. Chem. B* **110** 4527

[25]  Tikhonov A M 2006 *J. Chem. Phys.* **124** 164704

[26]  Luo G, Malkova S, Yoon J, Schultz D G, Lin B, Meron M, Benjamin I, Vanysek P and Schlossman M L 2006 *J. Electroanal. Chem.* **593** 142

[27]  Luo G, Malkova S, Yoon J, Schultz D G, Lin B, Meron M, Benjamin I, Vanysek P and Schlossman M L 2006 *Science* **311** 216

[28]  Tidswell I M, Ocko B M, Pershan P S, Wasserman S R, Whitesides G M and Axe J D 1990 *Phys. Rev. B* **41** 1111

[29]  Buff F P, Lovett R A and Stillinger F H 1965 *Phys. Rev. Lett.* **15** 621

[30]  Weeks J D 1977 *J. Chem. Phys.* **67** 3106

[31]  Sinha S K, Sirota E B, Garoff S and Stanley H B 1988 *Phys. Rev. B* **38** 2297

[32]  Braslau A, Deutsch M, Pershan P S, Weiss A H, Als-Nielsen J and Bohr J 1985 *Phys. Rev. Lett.* **54** 114

[33]  Braslau A, Pershan P S, Swislow G, Ocko B M and Als-Nielsen J 1988 *Phys. Rev. A* **38** 2457

[34]  Dietrich S, in: Domb C and Lebowitz J (eds.), Phase Transitions and Critical Phenomena, Academic Press, London, 1988, pp. 2

[35]  Bonn D and Ross D 2001 *Rep. Prog. Phys.* **64** 1085

**Table 1 Fitting Parameters for X-ray Reflectivity**

Fitting parameters for fits to the x-ray reflectivity data. Slabs are ordered water–1–2–3–hexane (or hexadecane); $L$ is the slab thickness; $L_{trans}$ is the calculated length of the all-trans alkanol ($L_{trans}$ = (m-1)x1.27Å (C—C) + 1.5Å (—CH$_3$) + 2.4Å (—CH$_2$OH)); $\rho$ is the normalized electron density; $\sigma$ is the interfacial roughness; $\sigma_{cap}$ is the roughness calculated from the measured interfacial tension using capillary wave theory. The electron densities are normalized to the value for bulk water (e.g., 0.3333 e$^-$/Å$^3$ at T = 25°C). The normalized hexane density is e.g., 0.692 at T = 20°C. The parameter $N$ is the total number of electrons per area in the interfacial region determined by the fitted electron density profile. Calculation of the area per molecule $A$ assumes that only alkanol molecules are in the interfacial region described by the slab model. C$_m$OH refers to the C$_m$-alkanol.

| System | Slab 1 $L_1$ (Å) | Slab 1 $\rho_1$ | Slab 2 $L_2$ (Å) | Slab 2 $\rho_2$ | Slab 3 $L_3$ (Å) | Slab 3 $\rho_3$ | $\sigma$ (Å) | $\sigma_{cap}$ (Å) | $L_{trans}$ (Å) | $N$ (e$^-$/Å$^2$) | $A$ (Å$^2$) |
|---|---|---|---|---|---|---|---|---|---|---|---|
| **Low Temperature:** | | | | | | | | | | | |
| *Hexadecane/Water Interface:* | | | | | | | | | | | |
| C$_{24}$OH (50.8°C) | 29$^{+1/-28}$ | 0.970$^{\pm 0.003}$ | 0.5$^{+28/-0.4}$ | 1.5$^{+0/-1}$ | 33$^{+1/-3}$ | 0.740$^{\pm 0.003}$ | 3.7$^{+0.1/-0.7}$ | 3.9$^{\pm 0.2}$ | 33.1 | 17.8$^{\pm 0.7}$ | 23$^{\pm 1}$ |
| C$_{30}$OH (24.9°C) | 9$^{\pm 6}$ | 1.17$^{+0.4/-0.1}$ | 0 | 0 | 18$^{\pm 2}$ | 0.89$^{\pm 0.01}$ | 4.5$^{\pm 1.5}$ | 3.9$^{\pm 0.2}$ | 40.7 | 8.9$^{+1.6/-1.1}$ | 28$^{\pm 4}$ |
| C$_{30}$OH (24.9°C) | 2$^{+6/-1}$ | 1.4$^{+0.2/-0.3}$ | 18$^{+1/-2}$ | 0.79$^{+0.02/-0.01}$ | 18 | 0.770$^{\pm 0.003}$ | 3.9 | 3.9$^{\pm 0.2}$ | 40.7 | 10$^{+2/-0.2}$ | 25$^{+0.3/-4}$ |
| *Hexane/Water Interface:* | | | | | | | | | | | |
| C$_{12}$OH (8.0°C) | 18$^{+1/-13}$ | 1.12$^{+0.4/-0.01}$ | 18$^{\pm 1}$ | 0.81$^{+0.03/-0.01}$ | 19$^{\pm 1}$ | 0.714$^{\pm 0.005}$ | 4.8$^{+0.1/-0.3}$ | 3.7$^{\pm 0.2}$ | 17.9 | 16$^{+1/-3}$ | 20$^{+4/-1}$ |
| C$_{24}$OH (21.9°C) | 5$^{+4/-3}$ | 1.24$^{+0.4/-0.1}$ | 10$^{+1/-1.5}$ | 0.95$^{+0.05/-0.03}$ | 14$^{\pm 1}$ | 0.81$^{\pm 0.01}$ | 3.3$^{+0.5/-1}$ | 4.5$^{\pm 0.2}$ | 33.1 | 9.0$^{+0.5/-0.4}$ | 22.4$^{\pm 1}$ |
| C$_{30}$OH (24.5°C) | 4$^{+5/-2}$ | 1.32$^{+0.3/-0.2}$ | 13$^{\pm 2}$ | 0.95$^{+0.02/-0.03}$ | 18$^{\pm 1}$ | 0.79$^{\pm 0.01}$ | 3.4$^{+0.4/-0.6}$ | 3.8$^{\pm 0.2}$ | 40.7 | 10.6$^{+0.5/-0.4}$ | 23.6$^{\pm 1}$ |
| **High Temperature:** | | | | | | | | | | | |
| *Hexadecane/Water Interface:* | | | | | | | | | | | |
| C$_{24}$OH (81.9°C) | | | | | | | 3.5$^{\pm 0.3}$ | 3.9$^{\pm 0.2}$ | | | |
| C$_{30}$OH (48.8°C) | | | | | | | 3.2$^{\pm 0.3}$ | 3.7$^{\pm 0.2}$ | | | |
| *Hexane/Water Interface:* | | | | | | | | | | | |
| C$_{12}$OH (55.0°C) | | | | | | | 4.5$^{\pm 0.2}$ | 3.8$^{\pm 0.2}$ | | | |
| C$_{24}$OH (45.3°C) | | | | | | | 5.0$^{\pm 0.2}$ | 3.5$^{\pm 0.2}$ | | | |
| C$_{30}$OH (45.0°C) | | | | | | | 4.8$^{\pm 0.3}$ | 3.8$^{\pm 0.2}$ | | | |

Figue1.  a) Slab model of x-ray reflectivity data for a monolayer of *n*-alkanol surfactant at the interface between water and a solution of alkanols in hexane or hexadecane.  Models in the text may vary in two ways from the illustration: the number of slabs can vary, and the slabs can represent multi-molecular layering, not just the monolayer illustrated.  *Z* labels the interface positions.  b) The kinematics of scattering in the right-handed rectangular coordinate system where the origin, *O,* is in the center of the x-ray footprint; here, the *xy* plane coincides with the water surface, the *y*-axis coincides with the projection of the incident beam's direction on the interface, and the *z*-axis is directed normal to the interface and opposite to the gravitational force. At the specular reflectivity condition, $\alpha = \beta$, and $\phi = 0$, $\alpha$ is the incident angle in the *yz* plane, $\beta$ is the angle in the vertical plane between the scattering direction and the interface, and $\phi$ is the angle in the *xy* plane between the incident beam's direction and the direction of the scattering. $k_{in}$ and $k_{sc}$ are, respectively, wave vectors of the incident beam and the beam scattered toward the point of observation. At the condition for specular reflectivity, the wave-vector transfer $q = k_{sc} - k_{in}$ has only one nonzero component, $q_z = (4\pi/\lambda)\sin\alpha$.

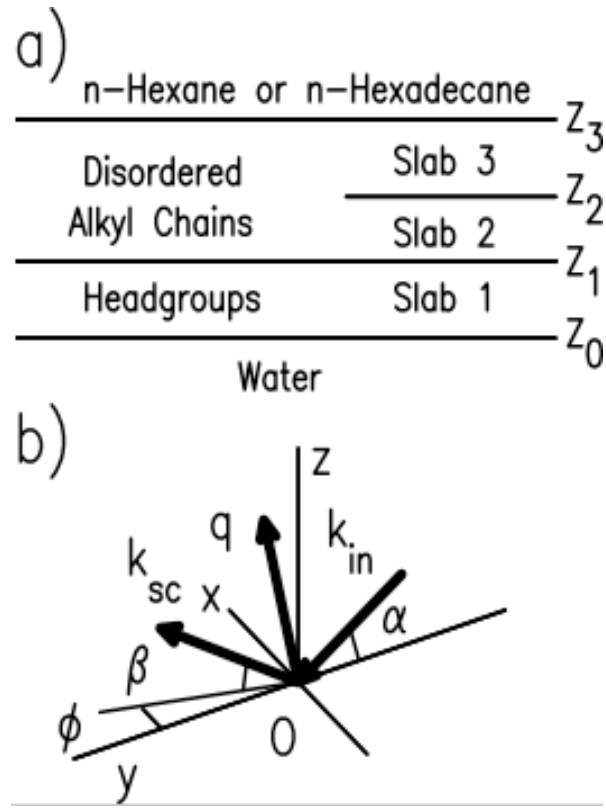

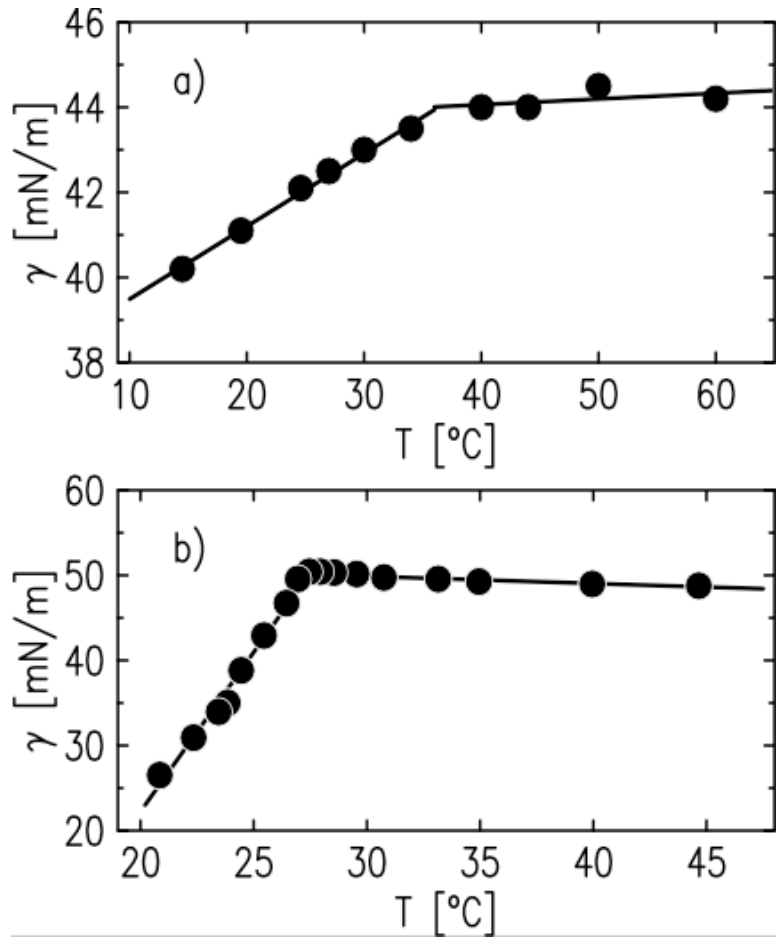

Figure 2 Temperature dependence of interfacial tension of the hexane/water interface: a) 45 mmol/kg $C_{12}$-alkanol solution in hexane; b) 0.7 mmol/kg $C_{30}$-alkanol solution in hexane [8].

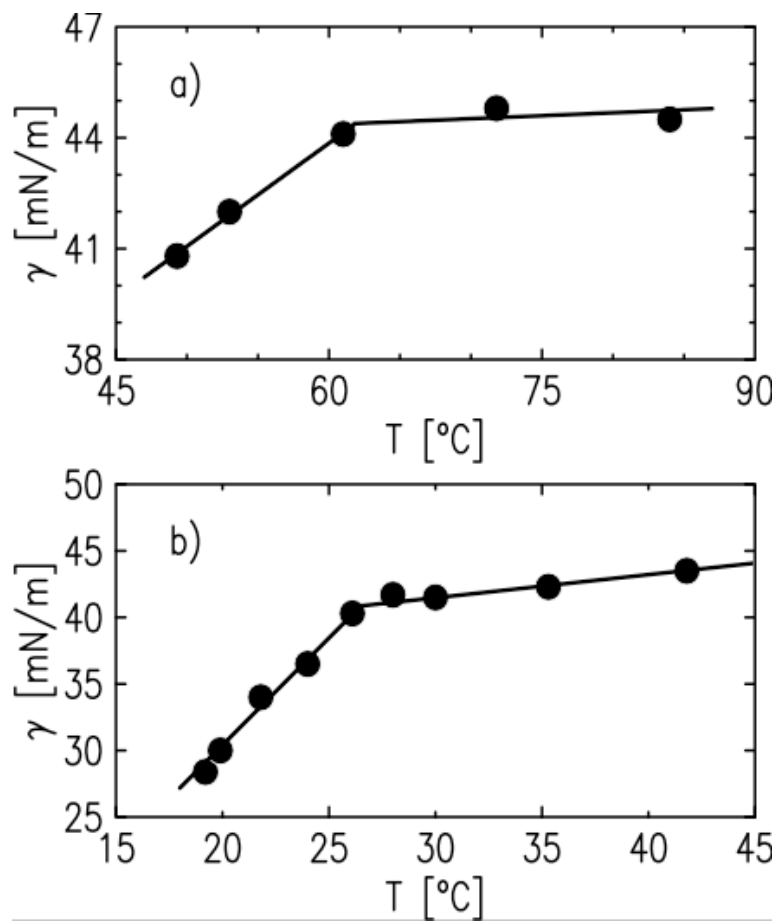

Figure 3 Temperature dependence of interfacial tension of the hexadecane/water interface: a) 4 mmol/kg $C_{24}$-alkanol solution in hexadecane; b) 0.2 mmol/kg $C_{30}$-alkanol solution in hexadecane.

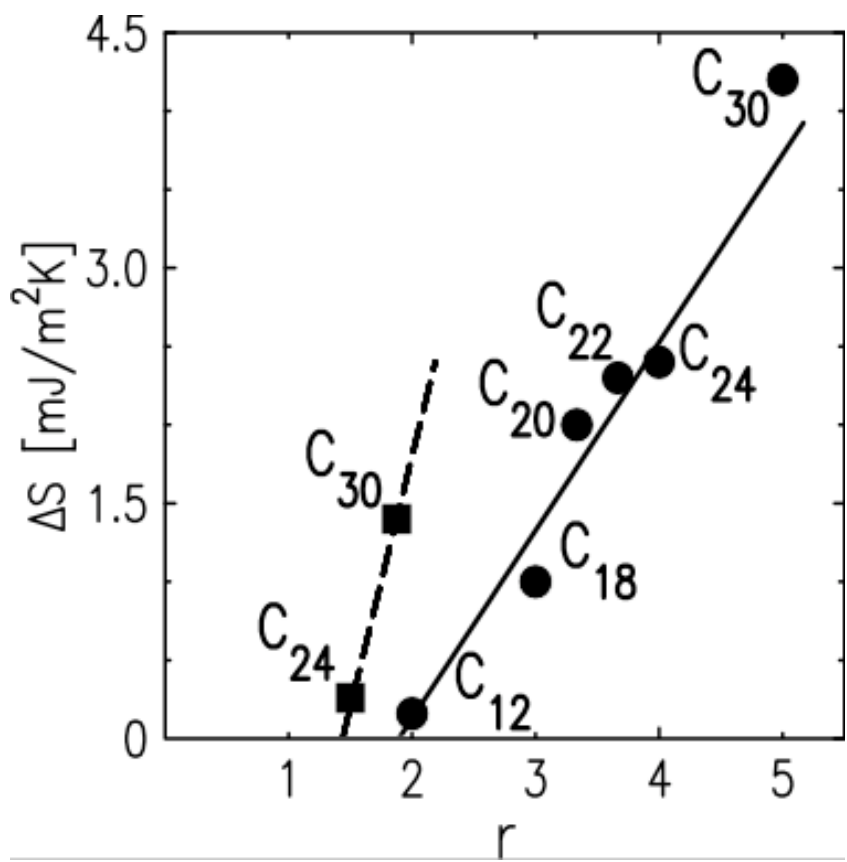

Figure 4 Change in interfacial excess entropy per unit area across the transition, $\Delta S_a^\sigma$ vs. $r$ for solutions in hexane (dots) and in hexadecane (squares), where $r = m/m_o$ is the ratio of the alkanol carbon number $m$ to the alkane solvent carbon number $m_o$. Symbols are labeled with the carbon number of the alkanol surfactant.

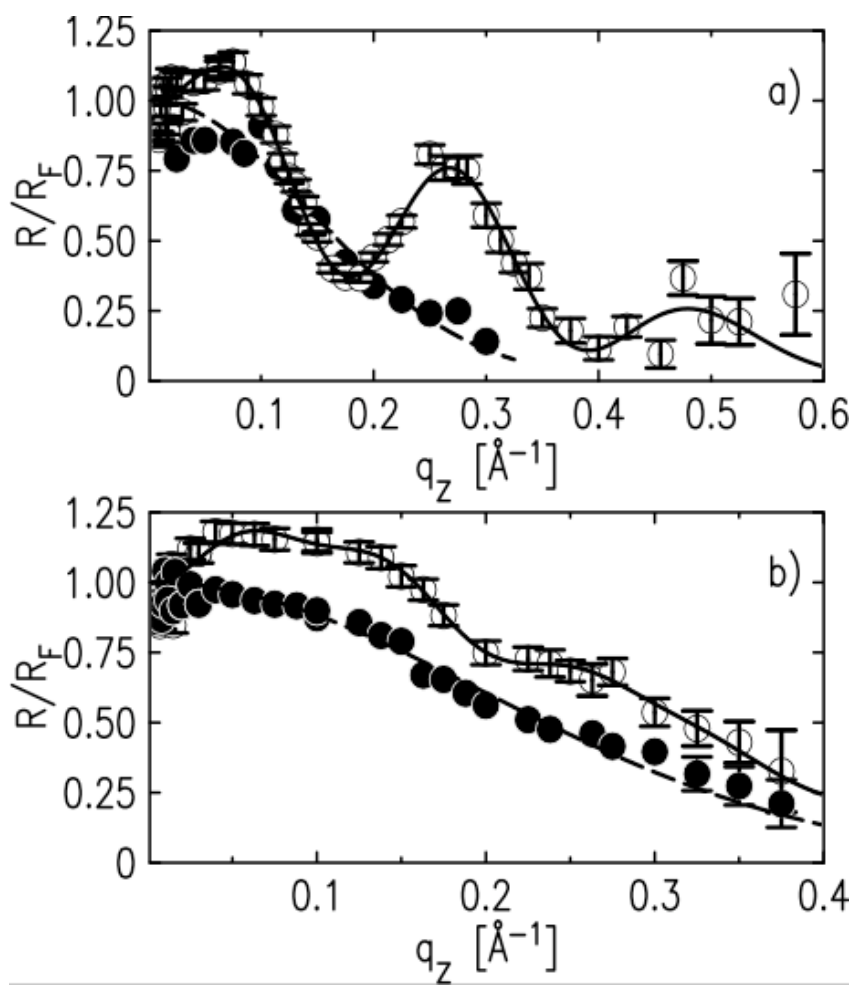

Figure 5 $C_{24}$-alkanol: x-ray reflectivity (normalized to the Fresnel reflectivity) as a function of the wave vector transfer normal to the interface. a) low temperature ($T = 21.9$ °C, circles) and high temperature ($T = 45.3$ °C, dots) measurements of $C_{24}$-alkanol at the hexane/water interface [8]. Solid line is a three slab model of a monolayer; dashed line is a one parameter fit. b) low temperature ($T = 50.8$ °C, circles) and high temperature ($T = 81.9$ °C, dots) measurements of $C_{24}$-alkanol at the hexadecane/water interface. Solid line is a two slab model of a bilayer; dashed line is a one parameter model.

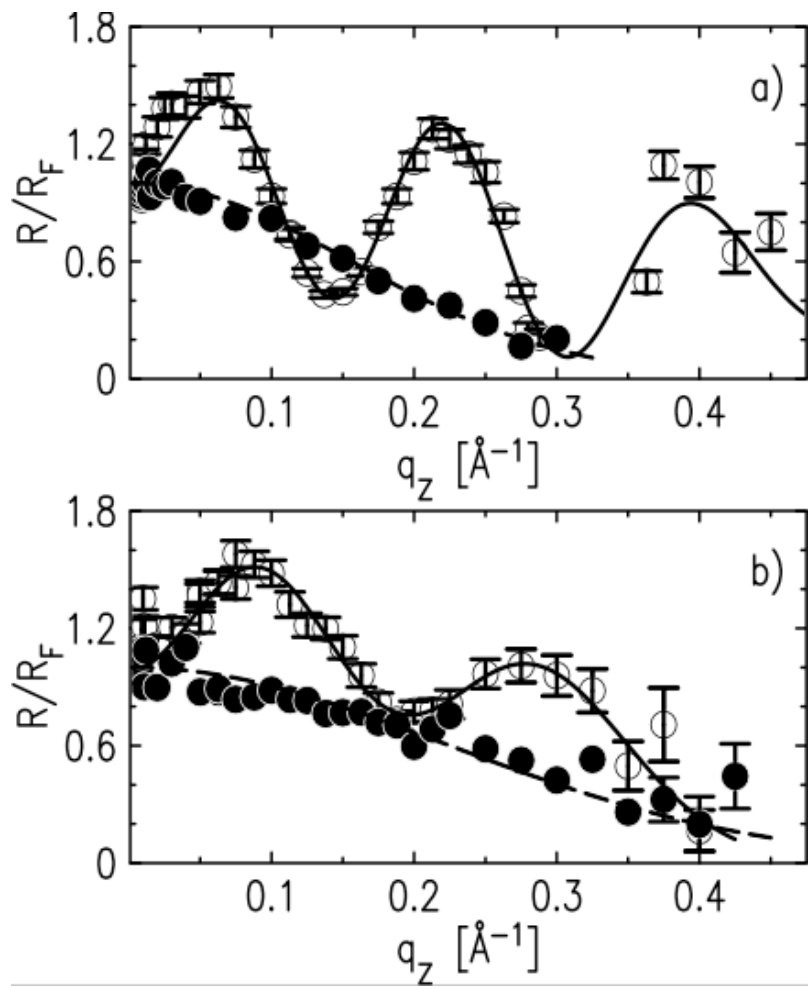

Figure 6 $C_{30}$-alkanol: x-ray reflectivity (normalized to the Fresnel reflectivity) as a function of the wave vector transfer normal to the interface. a) low temperature ($T = 24.5$ °C, circles) and high temperature ($T = 45.0$ °C, dots) measurements of $C_{30}$-alkanol at the hexane/water interface [8]. Solid line is a three slab model of a monolayer; dashed line is a one parameter model. b) low temperature ($T = 24.9$ °C, circles) and high temperature ($T = 48.8$ °C, dots) measurements of $C_{30}$-alkanol at the hexadecane/water interface. Solid line is a two slab model of a monolayer; dashed line is a one parameter model.

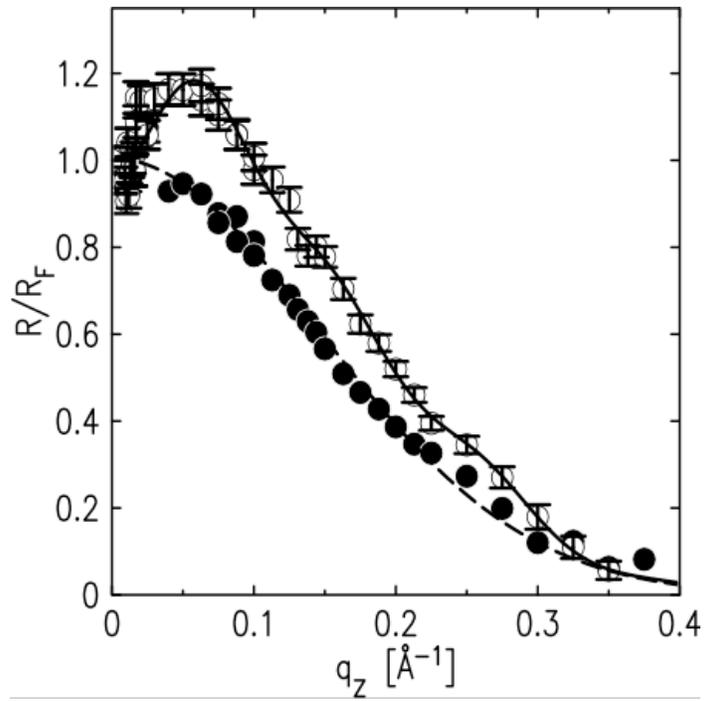

Figure 7  $C_{12}$-alkanol: x-ray reflectivity (normalized to the Fresnel reflectivity) as a function of the wave vector transfer normal to the interface.  Low temperature  $T = 8.0$ °C (circles) and high temperature $T = 55.0$ °C (dots) measurements of $C_{12}$-alkanol at the hexane/water interface.  Solid line is a three slab model of a tri-molecular layer;  dashed line is a one parameter model.

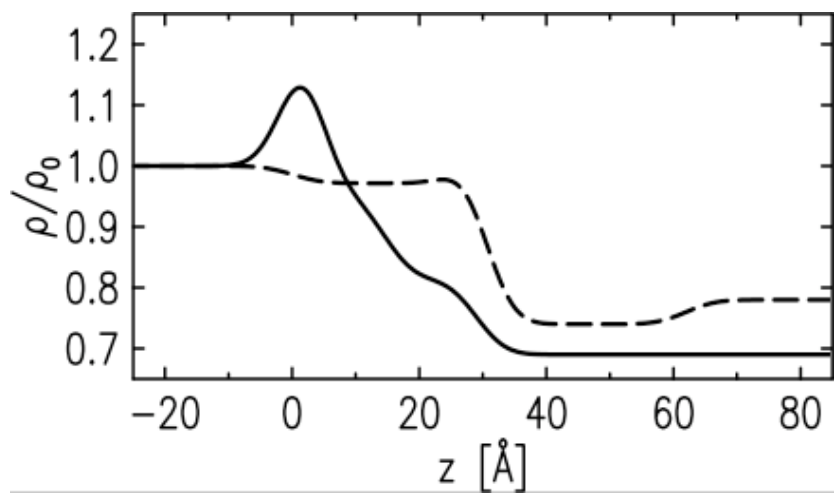

Figure 8  $C_{24}$-alkanol electron density profiles ($z < 0$ is bulk water) in the low temperature region: solid line is the three slab model of a monolayer at the hexane/water interface ($T = 21.9$ °C); dashed line is the three slab model of a molecular bilayer at the hexadecane/water interface ($T = 50.8$ °C).

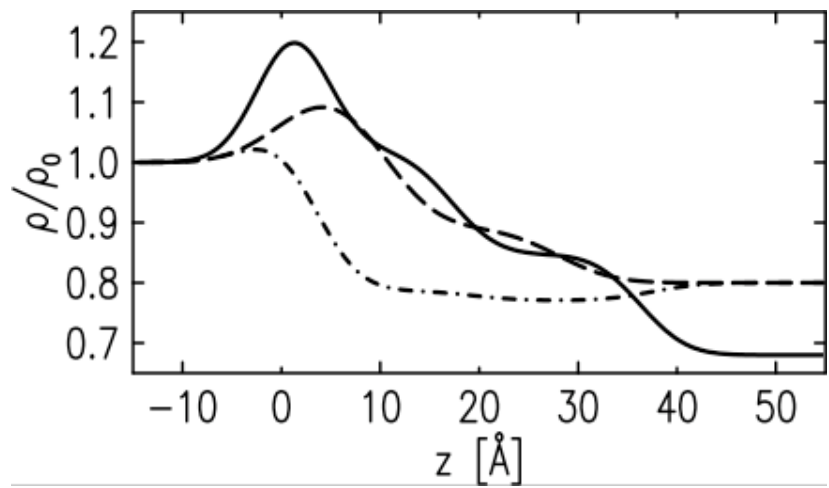

Figure 9  $C_{30}$-alkanol electron density profiles ($z < 0$ is bulk water) in the low temperature region: solid line is a three-slab model of a monolayer at the hexane/water interface ($T = 24.5$ °C); dashed line is a two slab model of a monolayer at the hexadecane/water interface ($T = 24.9$ °C); dash-dotted line is a three slab model of a monolayer at the hexadecane/water interface.

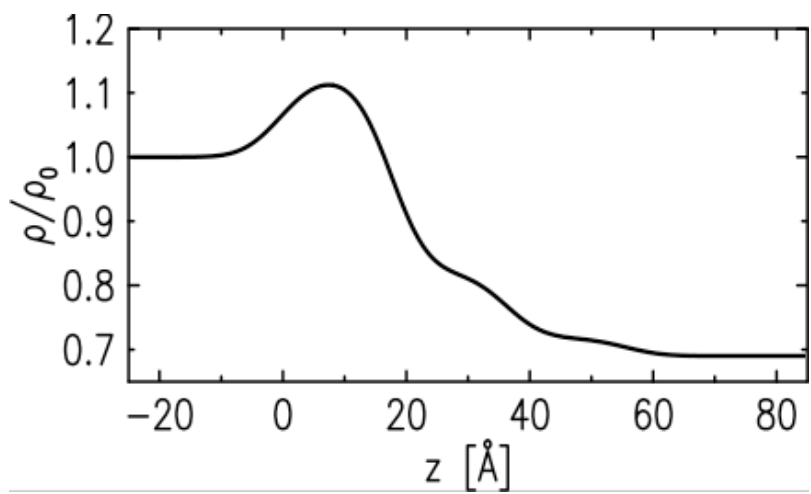

Figure 10  $C_{12}$-alkanol electron density profile ($z < 0$ is bulk water) in the low temperature region at $T = 8.0\ °C$ for a three slab model of a tri-molecular layer at the hexane/water interface.